\documentclass[prd,showpacs,notitlepage]{revtex4-1}
\usepackage{amsmath}
\usepackage{epsfig}
\usepackage{bm}
\usepackage{amssymb}
\usepackage[bookmarks=false]{hyperref}
\usepackage{graphicx}

\begin{document}

\title{Solving the heat-flow problem with transient relativistic fluid dynamics}
\author{G.\ S.\ Denicol${}^{a}$, H.\ Niemi${}^{b}$, I. Bouras${}^{a}$, E.\
Moln\'ar${}^{c,d}$, Z.\ Xu${}^{e}$, D.\ H.\ Rischke${}^{a,c}$, and C.\ Greiner${}^{a}$ }

\begin{abstract}
Israel-Stewart theory is a causal, stable formulation of relativistic
dissipative fluid dynamics. This theory has been shown to give a decent description of
the dynamical behavior of a relativistic fluid in cases where shear stress becomes
important. In principle, it should also be applicable to situations where
heat flow becomes important. However, it has been shown that there are
cases where Israel-Stewart theory cannot reproduce phenomena
associated with heat flow.
In this paper, we derive a relativistic dissipative fluid-dynamical
theory from kinetic theory which
provides a good description of all dissipative phenomena, including heat
flow. We explicitly demonstrate this
by comparing this theory with numerical solutions of the
relativistic Boltzmann equation.
\end{abstract}

\affiliation{$^{a}$Institut f\"ur Theoretische Physik, Johann Wolfgang
Goethe-Universit\"at, Max-von-Laue-Str.\ 1, D-60438 Frankfurt am Main,
Germany}

\affiliation{$^{b}$Department of Physics, P.O. Box 35 (YFL) FI-40014 University of
Jyv\"askyl\"a, Finland}

\affiliation{$^{c}$Frankfurt Institute for Advanced Studies, Ruth-Moufang-Str.\ 1,
D-60438 Frankfurt am Main, Germany}

\affiliation{$^{d}$ MTA-KFKI, Research Institute of Particle and
  Nuclear Physics, H-1525 Budapest, P.O.Box 49, Hungary}

\affiliation{$^{e}$Department of Physics, Tsinghua University, Beijing 100084, China}

\maketitle


\section{Introduction}

The derivation of a consistent theory of \textit{relativistic} fluid
dynamics has been a challenge for some time. The difficulty resides in the
parabolic nature of Navier-Stokes theory, which allows signals to propagate
with infinite speed. While in non-relativistic theories this unphysical
feature can be dismissed, in relativistic systems it leads to unstable
equations of motion and is simply unacceptable \cite{his}. A necessary (but
not sufficient) condition is that the equations of motion are hyperbolic.

For relativistic dilute gases, the Boltzmann equation provides a reasonable
description of the underlying microscopic theory and can serve as a starting
point to investigate the microscopic foundations of fluid dynamics. For the
sake of simplicity, in this paper we consider a single-component system of
massless and classical particles interacting only via elastic two-body
collisions with a constant cross section $\sigma $. The relativistic
Boltzmann equation for this type of system reads 
\begin{equation}
k^{\mu }\partial _{\mu }f_{\mathbf{k}}=\frac{1}{2}\int dK^{\prime
}dPdP^{\prime }\,W_{\mathbf{kk}\prime \rightarrow \mathbf{pp}\prime }\left(
f_{\mathbf{p}}f_{\mathbf{p}^{\prime }}-f_{\mathbf{k}}f_{\mathbf{k}^{\prime
}}\right)\; ,  \label{BoltzmannEq}
\end{equation}%
where $f_{\mathbf{k}}$ is the single-particle distribution function, $W_{%
\mathbf{kk}\prime \rightarrow \mathbf{pp}\prime }$ is the Lorentz-invariant
transition rate, and $dK\equiv \,gd^{3}\mathbf{k/}\left[ (2\pi )^{3}k^{0}%
\right] $ is the Lorentz-invariant momentum-space volume, with $g$ being the
number of internal degrees of freedom. The right-hand side of the Boltzmann
equation describes the change of the single-particle distribution function
due to elastic collisions of two particles with incoming momenta $k$ and $%
k^{\prime }$, and outgoing momenta $p$ and $p^{\prime }$.

The conservation laws follow from the first two moments of the Boltzmann
equation as a consequence of the conservation of particle number and
energy-momentum in individual collisions, $\partial _{\mu }N^{\mu }=0,\
\partial _{\mu }T^{\mu \nu }=0$. The particle four-current, $N^{\mu }$, and
the energy-momentum tensor, $T^{\mu \nu }$, can be cast in the usual form,%
\begin{equation}
N^{\mu }=nu^{\mu }+n^{\mu },\text{ \ }T^{\mu \nu }=\varepsilon u^{\mu
}u^{\nu }-P_{0}\Delta ^{\mu \nu }+\pi ^{\mu \nu },  \label{N_mu}
\end{equation}%
where $n$ is the particle number density, $n^{\mu }$ is the particle
diffusion four-current, $\varepsilon $ is the energy density, $u^{\mu }$ is
the fluid four-velocity defined in the Landau frame \cite{Landau}, i.e., $%
u_{\nu }T^{\mu \nu }=\varepsilon u^{\mu }$, $P_{0}$ is the thermodynamic
pressure, and $\pi ^{\mu \nu }$ is the shear-stress tensor. Since we consider
a massless gas, the bulk viscous pressure is always zero. The additional
equations needed in order to solve the conservation laws, i.e., the
equations of motion for $n^{\mu }$ and $\pi ^{\mu \nu }$, must be derived by
properly matching fluid dynamics to the relativistic Boltzmann equation.

In the framework of relativistic kinetic theory, Israel and Stewart were
among the first to derive a relativistic theory of fluid dynamics that is
causal and stable \cite{IS}. In the Israel-Stewart (IS) formulation, the
single-particle distribution function is expanded around its local
equilibrium value in terms of a series of Lorentz tensors formed of particle
four-momentum $k^{\mu }$, i.e., $1,\,k^{\mu },\,k^{\mu }k^{\nu },\,\ldots $.
In order to derive fluid dynamics, this series is truncated at second order
in momentum, i.e., one only keeps the tensors $1,\,k^{\mu }$, and $k^{\mu
}k^{\nu }$, the so-called 14-moment approximation. The 14 coefficients of
this truncated expansion are uniquely matched to the 14 components of $%
N^{\mu }$ and $T^{\mu \nu }$. Finally, the equations of motion for $n^{\mu }$
and $\pi ^{\mu \nu }$ are obtained by substituting the truncated moment
expansion into the second moment of the Boltzmann equation.

In the past five years, IS theory has been widely applied to
ultrarelativistic heavy-ion collisions in order to describe the time
evolution of the quark-gluon plasma (QGP) and the freeze-out of the hadron
resonance gas appearing in the late stages of the collision. However, in
heavy-ion collisions extreme conditions occur which question the validity of
fluid dynamics. The QGP created at the Relativistic Heavy Ion Collider
(RHIC) and, recently, at the Large Hadron Collider (LHC) is not only the
fluid with the smallest space-time extension ($\sim 10$ fm) ever created in
nature but also the one where the space-time gradients of the fluid fields,
for instance energy density $\varepsilon $, are the largest ($\sim |\partial
_{\mu }\varepsilon |/\varepsilon \sim 1/$fm) ever encountered. On the other
hand, Israel and Stewart's derivation lacks a small parameter, such as the
Knudsen number, with which one can do power counting and systematically
improve the approximation to describe higher-order gradients. Thus, the
applicability of IS theory to the extreme conditions reached in heavy-ion
collisions is, at the very least, not clear.

One way to investigate the applicability of IS theory is to compare the
solutions of this theory with numerical solutions of the Boltzmann equation 
\cite{Pasi,Bouras_prl, Bouras, el, dkr}. So far, it was confirmed that, at least for
some special problems, the IS equations \cite{IS} are not in good agreement
with the numerical solution of the Boltzmann equation. In these cases, it was
shown by some of the present authors \cite{Bouras} that the IS formalism seems to be
unable to describe heat flow, even when the Knudsen number is very small. In
this paper, we would like to explain the reason for this discrepancy and
propose a solution which works at least if the Knudsen is not too large.

Recently, a systematic derivation of fluid dynamics from the Boltzmann
equation was introduced in Ref.\ \cite{DNMR}. The main difference between IS
theory and the theory derived in Ref.\ \cite{DNMR} is that the latter does
not truncate the moment expansion of the single-particle distribution
function. Instead, dynamical equations for all its moments are considered
and solved by separating the slowest microscopic time scale from the faster
ones. Then, the resulting fluid-dynamical equations are truncated according
to a systematic power-counting scheme using the inverse Reynolds number $%
\mathrm{R}^{-1}\sim \left\vert n^{\mu }\right\vert /n\sim \left\vert \pi
^{\mu \nu }\right\vert /P_{0}$ and the Knudsen number $\mathrm{Kn}=\lambda _{%
\mathrm{mfp}}/L$, with $\lambda _{\mathrm{mfp}}$ being the mean free-path
and $L$ a characteristic macroscopic distance scale, e.g. $L^{-1}\sim
\partial _{\mu }u^{\mu }$. The values of the transport coefficients of fluid
dynamics are obtained by resumming the contributions from all moments of the
single-particle distribution function, similar to what happens in the
Chapman-Enskog expansion \cite{CE}.

In this paper, we review the basic aspects of this resummed transient
relativistic fluid-dynamical theory (RTRFD). We explicitly derive the
equations of motion of RTRFD including terms up to second order in the
Knudsen number. We then demonstrate that this method is also able to handle
problems with strong initial gradients in pressure or particle number
density. This resolves the previously observed differences between the
solution of IS theory and of the Boltzmann equation \cite{Bouras} mentioned
above. We conclude that these differences were caused by the uncontrolled
truncation procedure of the expansion of the single-particle distribution
function in terms of Lorentz tensors in four-momentum as employed in IS
theory.

This paper is organized as follows. In Sec.\ \ref{RTRFD}, we review the
basic aspects of RTRFD. One important point is that the fluid-dynamical
equations of motion, as derived in Ref.\ \cite{DNMR}, become parabolic once
second-order terms in Knudsen number are included. In Sec.\ \ref{Hyper}, we
explain how to obtain hyperbolic equations of motion in the RTRFD formalism
up to second order in the Knudsen number. In Sec.\ \ref{Comp}, we compare
solutions obtained within RTRFD for various levels of approximation and
within IS theory with numerical solutions of the Boltzmann equation computed
using BAMPS \cite{BAMPS}. In Sec. \ref{Concl} we discuss our results and
draw conclusions. The Appendix contains intermediate steps of our
calculations. We use natural units $\hbar =c=k_{B}=1$. The metric tensor is $%
g_{\mu \nu }=\mathrm{diag}\,(+,-,-,-)$.

\section{Review of resummed transient relativistic fluid dynamics}

\label{RTRFD}

In RTRFD \cite{DNMR}, $f_{\mathbf{k}}$ is expanded in terms of an
orthonormal and complete basis in momentum space. The expansion basis
contains two basic ingredients: The first are the \textit{irreducible}
tensors, $1,$ $k^{\left\langle \mu \right\rangle },$ $k^{\left\langle \mu
_{1}\right. }k^{\left. \mu _{2}\right\rangle },\ldots ,$ $k^{\left\langle
\mu _{1}\right. }\cdots k^{\left. \mu _{m}\right\rangle }$, which form a 
\textit{complete and orthogonal\/} set, analogously to the spherical
harmonics \cite{DNMR, DeGroot, Anderson}. Here, we use the notation $%
A^{\left\langle \mu _{1}\cdots \mu _{\ell }\right\rangle }$ $\equiv \Delta
_{\nu _{1} \cdots\nu _{\ell }}^{\mu _{1}\cdots\mu _{\ell }} A^{\nu
_{1}\cdots\nu _{\ell }}$, with $\Delta _{\nu _{1}\cdots \nu _{m}}^{\mu
_{1}\cdots \mu _{m}}$. The latter quantities are projectors onto the subspaces
orthogonal to $u^\mu$. Their definition is explicitly given in Refs.\ \cite%
{DNMR,DeGroot}. Except for $m=1$, where $\Delta^\mu_\nu= g^\mu_\nu - u^\mu
u_\nu$, they are traceless. E.g., for $m=2$, $\Delta^{\mu \nu}_{\alpha
\beta} = (\Delta^\mu_\alpha \Delta^\nu_\beta + \Delta^\mu_\beta
\Delta^\nu_\alpha)/2 - \Delta^{\mu \nu} \Delta_{\alpha \beta}/3$. Note that
the expansion of $f_{\mathbf{k}}$ in IS theory is not in terms of the
irreducible tensors $k^{\left\langle \mu _{1}\right. }\cdots k^{\left. \mu
_{m}\right\rangle }$, but in terms of the tensors $k^{\mu_1} \cdots
k^{\mu_m} $ which are complete but neither irreducible nor orthogonal.

The second ingredient are orthogonal polynomials in $E_{\mathbf{k}}=u^{\mu
}k_{\mu }$, $P_{n\mathbf{k}}^{\left( \ell \right)
}=\sum_{r=0}^{n}a_{nr}^{\left( \ell \right) }E_{\mathbf{k}}^{r}$. For
details in constructing the polynomials, see Ref.\ \cite{DNMR}. Then, $f_{%
\mathbf{k}}$ is expanded as 
\begin{equation}
f_{\mathbf{k}}=f_{0\mathbf{k}}+f_{0\mathbf{k}}\sum_{\ell =0}^{\infty
}\sum_{n=0}^{N_{\ell }}\mathcal{H}_{n\mathbf{k}}^{\left( \ell \right)
}\;\rho _{n}^{\mu _{1}\cdots \mu _{\ell }}\;k_{\left\langle \mu _{1}\right.
}\cdots k_{\left. \mu _{\ell }\right\rangle }\;,  \label{fexpansion2}
\end{equation}%
where $f_{0\mathbf{k}}=\exp \left( \alpha _{0}-\beta _{0}E_{\mathbf{k}%
}\right) $ is the local equilibrium distribution function, with $\alpha
_{0}=\mu /T$ being the ratio of chemical potential to temperature and $\beta
_{0}=1/T$ the inverse temperature. We further introduced the
energy-dependent coefficients $\mathcal{H}_{n\mathbf{k}}^{(\ell )}\equiv
\left( W^{(\ell )}/\ell !\right) \sum_{m=n}^{N_{\ell }}a_{mn}^{\left( \ell
\right) }P_{m\mathbf{k}}^{(\ell )}$, with a normalization constant $W^{(\ell
)}$, and the irreducible moments of $\delta f_{\mathbf{k}}=f_{\mathbf{k}%
}-f_{0\mathbf{k}}$, 
\begin{equation}
\rho _{r}^{\mu _{1}\cdots \mu _{\ell }}\equiv \int dK\text{ }E_{\mathbf{k}%
}^{r}\;k^{\left\langle \mu _{1}\right. }\cdots k^{\left. \mu _{\ell
}\right\rangle }\text{ }\delta f_{\mathbf{k}}.  \label{rho}
\end{equation}%
Some of the irreducible moments are related to
the fields in Eq.\ (\ref{N_mu}): $n^{\mu }=\rho _{0}^{\mu }$ and $\pi ^{\mu
\nu }=\rho _{0}^{\mu \nu }$. The values of $\alpha _{0}$ and $\beta _{0}$
are defined by the matching conditions, $n\equiv \langle E_{\mathbf{k}%
}\rangle _{0},\;\;\;\varepsilon \equiv \left\langle E_{\mathbf{k}%
}^{2}\right\rangle _{0}$, where $\langle \cdots \rangle _{0}\equiv \int
dK\left( \cdots \right) f_{0\mathbf{k}}$. The matching conditions and the
definition of $u^{\mu }$ according with the Landau picture imply that the
following moments should vanish: $\rho _{1}=\rho _{2}=\rho _{1}^{\mu }=0$.

The equations of motion for $\rho _{r}^{\mu }$ and $\rho _{r}^{\mu \nu }$
together with their respective transport coefficients were derived in Ref. 
\cite{DNMR}. Since we are investigating the massless limit, the scalar
moments $\rho _{r}$ will not play a dominant role (they contribute mainly to
the bulk viscous pressure) and we set them to zero. We also neglect all
irreducible moments with tensor rank higher than $2$, since they are
traditionally not considered in fluid dynamics. The role of such moments
will be investigated in a future work. Then, the equations for the first-
and second-rank tensors $\rho _{r}^{\mu }$ and $\rho _{r}^{\mu \nu }$ read 
\begin{align}
\dot{\rho}_{r}^{\left\langle \mu \right\rangle }+\sum_{n=0,\neq 1}^{N_{\ell
}}\mathcal{A}_{rn}^{(1)}\rho _{n}^{\mu }& =\alpha _{r}^{(1)}I^{\mu }+\rho
_{r}^{\nu }\omega _{\left. {}\right. \nu }^{\mu }-\frac{r+3}{3}
\, \rho _{r}^{\mu }\theta -\Delta _{\lambda }^{\mu }\nabla _{\nu
}\rho _{r-1}^{\lambda \nu }+r\rho _{r-1}^{\mu \nu }\dot{u}_{\nu }-\frac{2r+3}{5}%
\, \rho _{r}^{\nu }\sigma _{\nu }^{\mu }+\frac{\beta
_{0}I_{r+2,1}}{\varepsilon _{0}+P_{0}}\Delta _{\nu }^{\mu }\partial
_{\lambda }\pi ^{\lambda \nu },  \notag \\
\dot{\rho}_{r}^{\left\langle \mu \nu \right\rangle }+\sum_{n=0}^{N_{\ell }}%
\mathcal{A}_{rn}^{(2)}\rho _{n}^{\mu \nu }& =2\alpha _{r}^{(2)}\sigma ^{\mu
\nu }-\frac{2}{7}\left( 2r+5\right) \rho _{r}^{\lambda \left\langle \mu
\right. }\sigma _{\lambda }^{\left. \nu \right\rangle }+2\rho _{r}^{\lambda
\left\langle \mu \right. }\omega _{\left. {}\right. \lambda }^{\left. \nu
\right\rangle }+\frac{2}{5}\nabla ^{\left\langle \mu \right. }\rho
_{r+1}^{\left. \nu \right\rangle }-\frac{2}{5}\left( r+5\right) \rho
_{r+1}^{\left\langle \mu \right. }\dot{u}^{\left. \nu \right\rangle }-\frac{r+4%
}{3}\, \rho _{r}^{\mu \nu }\theta ,  \label{motion}
\end{align}%
where $\dot{\rho}_{r}^{\left\langle \mu _{1}\cdots \mu _{\ell }\right\rangle
}\equiv \Delta _{\nu _{1}\cdots \nu _{\ell }}^{\mu _{1}\cdots \mu _{\ell
}}u^{\mu }\partial _{\mu }\rho _{r}^{\nu _{1}\cdots \nu _{\ell }}$, $I^{\mu
}=\nabla ^{\mu }\alpha _{0}$, $\sigma ^{\mu \nu }=\partial ^{\left\langle
\mu \right. }u^{\left. \nu \right\rangle }$, and $\nabla ^{\mu }=\partial
^{\left\langle \mu \right\rangle }$ \cite{DNMR}. We also defined the
thermodynamic integrals 
\begin{equation}
I_{nq}\left( \alpha _{0},\beta _{0}\right) =\frac{1}{\left( 2q+1\right) !!}%
\int dKE_{\mathbf{k}}^{n-2q}\left( -\Delta ^{\alpha \beta }k_{\alpha
}k_{\beta }\right) ^{q}f_{0\mathbf{k}}\;.  \label{Jnq}
\end{equation}%
The coefficients 
\begin{align}
\mathcal{A}_{rn}^{\left( \ell \right) }& =\frac{1}{4\ell +2}\int
dKdK^{\prime }dPdP^{\prime }W_{\mathbf{kk}\prime \rightarrow \mathbf{pp}%
\prime }f_{0\mathbf{k}}f_{0\mathbf{k}\prime }\tilde{f}_{0\mathbf{p}}\tilde{f}%
_{0\mathbf{p}\prime }E_{\mathbf{k}}^{r-1}k^{\left\langle \nu _{1}\right.
}\cdots k^{\left. \nu _{\ell }\right\rangle }  \notag \\
& \times \left( \mathcal{H}_{\mathbf{k}n}^{\left( \ell \right)
}k_{\left\langle \nu _{1}\right. }\cdots k_{\left. \nu _{\ell }\right\rangle
}+\mathcal{H}_{\mathbf{k}^{\prime }n}^{\left( \ell \right) }k_{\left\langle
\nu _{1}\right. }^{\prime }\cdots k_{\left. \nu _{\ell }\right\rangle
}^{\prime }-\mathcal{H}_{\mathbf{p}n}^{\left( \ell \right) }p_{\left\langle
\nu _{1}\right. }\cdots p_{\left. \nu_\ell \right\rangle }-\mathcal{H}_{%
\mathbf{p}^{\prime }n}^{\left( \ell \right) }p_{\left\langle \nu _{1}\right.
}^{\prime }\cdots p_{\left. \nu _{\ell }\right\rangle }^{\prime }\right)
\label{integrals}
\end{align}%
contain all the information of the microscopic theory, while $\alpha
_{r}^{(\ell )}$ are complicated functions of $\beta _{0}$ and $\alpha _{0}$ 
\cite{DNMR}.

We now identify the microscopic time scales that dominate the long-time
dynamics of the Boltzmann equation. This can be achieved by finding the
normal modes of Eqs.\ (\ref{motion}), i.e., by diagonalizing the matrices $%
\mathcal{A}^{\left( \ell \right) }$. Then, one obtains the following (in
the linear regime, decoupled) set of equations of motion,
\begin{eqnarray}
\dot{X}_{r}^{\left\langle \mu \right\rangle }+\chi _{r}^{(1)}X_{r}^{\mu }
&=&\beta _{r}^{(1)}I^{\mu }+\mathcal{X}_{r}^{\mu },  \notag \\
\dot{X}_{r}^{\left\langle \mu \nu \right\rangle }+\chi _{r}^{(2)}X_{r}^{\mu
\nu } &=&\beta _{r}^{(2)}\sigma ^{\mu \nu }+\mathcal{X}_{r}^{\mu \nu },
\label{help formulas}
\end{eqnarray}%
where $\chi _{r}^{(\ell )}$ are the eigenvalues of $\mathcal{A}^{\left( \ell
\right) }$ and $X_{r}^{\mu _{1}\cdots \mu _{\ell }}\equiv
\sum_{j=0}^{N_{\ell }}\left( \Omega ^{-1}\right) _{rm}^{\left( \ell \right)
}\rho _{m}^{\mu _{1}\cdots \mu _{\ell }}$ are the eigenmodes of the
linearized Boltzmann equation, with $\left( \Omega ^{-1}\right) ^{\left(
\ell \right) }\mathcal{A}^{\left( \ell \right) }\Omega ^{\left( \ell \right)
}=\mathrm{diag}(\chi _{0}^{(\ell )},\ldots ,\chi _{r}^{(\ell )},\ldots )$.
The terms $\mathcal{X}_{r}^{\mu }$ and $\mathcal{X}_{r}^{\mu \nu }$
represent nonlinear terms and terms containing derivatives of the
eigenmodes. The coefficients $\beta _{r}^{(\ell )}$ are complicated
functions of $\alpha _{0}$ and $\beta _{0}$ \cite{DNMR}. Without loss of
generality, we order the eigenmodes of the linearized Boltzmann equation, $%
X_{r}^{\mu _{1}\cdots \mu _{\ell }}$, according to increasing $\chi
_{r}^{(\ell )}$, i.e., in such a way that $\chi _{r}^{(\ell )}<\chi
_{r+1}^{(\ell )}$, $\forall $ $\ell $.

It is clear that when the terms $\mathcal{X}_{r}^{\mu }$ and $\mathcal{X}%
_{r}^{\mu \nu }$in Eqs.\ (\ref{help formulas}) are small, each eigenmode
relaxes independently to its asymptotic solution on time scales given by $%
1/\chi _{r}^{(\ell )}$. We shall refer to the solution at asymptotically
long times as Navier-Stokes value. The slowest varying eigenmodes of the
Boltzmann equation are those for $r=0$. At very long times, only the
transient dynamics of these eigenmodes has to be resolved, while all quickly
varying modes are assumed to have already relaxed to their corresponding
Navier-Stokes values. In practice, this means we have to solve Eqs.\ (\ref%
{help formulas}) for $r=0$ dynamically, 
\begin{eqnarray}
\dot{X}_{0}^{\left\langle \mu \right\rangle }+\chi _{0}^{(1)}X_{0}^{\mu }
&=&\beta _{0}^{(1)}I^{\mu }+\mathcal{X}_{0}^{\mu },  \notag \\
\dot{X}_{0}^{\left\langle \mu \nu \right\rangle }+\chi _{0}^{(2)}X_{0}^{\mu
\nu } &=&\beta _{0}^{(2)}\sigma ^{\mu \nu }+\mathcal{X}_{0}^{\mu \nu },
\end{eqnarray}%
while all other eigenmodes ($r\geq 1$) are approximated to be very close to
their asymptotic (i.e., Navier-Stokes) values,%
\begin{equation}
X_{r}^{\mu }=\frac{\beta _{r}^{(1)}}{\chi _{r}^{(1)}}I^{\mu }+\ldots ,\text{
\ }X_{r}^{\mu \nu }=\frac{\beta _{r}^{(2)}}{\chi _{r}^{(2)}}\sigma ^{\mu \nu
}+\ldots ,\text{ \ }r\geq 1\text{ ,}  \label{asymptotic}
\end{equation}%
where the dots indicate deviations from the asymptotic solution which are at
least of order $\mathcal{O}(\mathrm{Kn}^{2})$. One observes that the
eigenmodes of index $r\geq 1$ are all of first order in Knudsen number. In
the asymptotic regime and to first order in the Knudsen number, $I^{\mu }$, $%
\sigma ^{\mu \nu }$ on the r.h.s.\ of the relations (\ref{asymptotic}) can
also be expressed in terms of any other eigenmode, e.g.\ the $m$th one, $%
m\geq 1$, such that for all $n\geq 1,\,m\neq n$, 
\begin{eqnarray}
X_{n}^{\mu } &=&\frac{\chi _{m}^{(1)}}{\chi _{n}^{(1)}}\frac{\beta _{n}^{(1)}%
}{\beta _{m}^{(1)}}X_{m}^{\mu }+\mathcal{O}(\mathrm{Kn}^{2},\mathrm{Kn\,R}%
^{-1},\mathrm{R}^{-2})\text{ }, \\
X_{n}^{\mu \nu } &=&\frac{\chi _{m}^{(2)}}{\chi _{n}^{(2)}}\frac{\beta
_{n}^{(2)}}{\beta _{m}^{(2)}}X_{m}^{\mu \nu }+\mathcal{O}(\mathrm{Kn}^{2},%
\mathrm{Kn\,R}^{-1},\mathrm{R}^{-2})\text{ }.
\end{eqnarray}%
Then, choosing, e.g. $m=2$ for $\ell =1$ and $m=1$ for $\ell =2$, in the
above relations and using $\rho _{r}^{\mu _{1}\cdots \mu _{\ell }}\equiv
\sum_{n=0}^{N_{\ell }}\Omega _{rn}^{\left( \ell \right) }X_{n}^{\mu
_{1}\cdots \mu _{\ell }}$, we can write,%
\begin{eqnarray}
\rho _{r}^{\mu } &=&\Omega _{r0}^{\left( 1\right) }X_{0}^{\mu }+\frac{\chi
_{2}^{(1)}}{\beta _{2}^{(1)}}\left[ \sum_{n=2}^{N_{1}}\Omega _{rn}^{\left(
1\right) }\frac{\beta _{n}^{(1)}}{\chi _{n}^{(1)}}\right] X_{2}^{\mu
}=\Omega _{r0}^{\left( 1\right) }X_{0}^{\mu }+\mathcal{O}(\mathrm{Kn})\text{ 
},  \notag \\
\rho _{r}^{\mu \nu } &=&\Omega _{r0}^{\left( 2\right) }X_{0}^{\mu \nu }+%
\frac{\chi _{1}^{(2)}}{\beta _{1}^{(2)}}\left[ \sum_{n=1}^{N_{2}}\Omega
_{rn}^{\left( 2\right) }\frac{\beta _{n}^{(2)}}{\chi _{n}^{(2)}}\right]
X_{1}^{\mu \nu }=\Omega _{r0}^{\left( 2\right) }X_{0}^{\mu \nu }+\mathcal{O}(%
\mathrm{Kn})\text{ }.  \label{flamengo2}
\end{eqnarray}%
Using $\rho _{0}^{\mu }=n^{\mu }$ and $\rho _{0}^{\mu \nu }=\pi ^{\mu \nu }$%
, we replace the zeroth eigenmodes, $X_{0}^{\mu }$ and $X_{0}^{\mu \nu }$,
in Eq.~(\ref{flamengo2}) and express all irreducible moments, up to first
order in Knudsen number, in terms of $n^{\mu }$ and $\pi ^{\mu \nu }$,%
\begin{eqnarray}
\rho _{r}^{\mu } &=&\Omega _{r0}^{\left( 1\right) }n^{\mu }+\frac{\chi
_{2}^{(1)}}{\beta _{2}^{(1)}}\left( \kappa _{r}-\Omega _{r0}^{\left(
1\right) }\kappa _{0}\right) X_{2}^{\mu }=\Omega _{r0}^{\left( 1\right)
}n^{\mu }+\mathcal{O}(\mathrm{Kn})\text{ },  \notag \\
\rho _{r}^{\mu \nu } &=&\Omega _{r0}^{\left( 2\right) }\pi ^{\mu \nu }+2\,%
\frac{\chi _{1}^{(2)}}{\beta _{1}^{(2)}}\left( \eta _{r}-\Omega
_{r0}^{\left( 2\right) }\eta _{0}\right) X_{1}^{\mu \nu }=\Omega
_{r0}^{\left( 2\right) }\pi ^{\mu \nu }+\mathcal{O}(\mathrm{Kn})\text{ },
\label{noname}
\end{eqnarray}%
where $\kappa _{r}=\sum_{k=0,\neq 1}^{N_{1}}\tau _{rk}^{\left( 1\right)
}\alpha _{k}^{\left( 1\right) }$, $\eta _{r}=$ $\sum_{k=0}^{N_{2}}\tau
_{rk}^{\left( 2\right) }\alpha _{k}^{\left( 2\right) }$, with $\tau ^{(\ell
)}=(\mathcal{A}^{-1})^{\ell }$, and we set $\Omega _{00}^{\left( \ell
\right) }=1$. This formula is exact even for $r=0$, for which the
first-order terms in Knudsen number vanish.

In Ref.\ \cite{DNMR}, $X_{2}^{\mu }$ and $X_{1}^{\mu \nu }$ in Eqs.\
(\ref{noname}) are substituted
 by their Navier-Stokes values, i.e., 
\begin{equation}
X_{2}^{\mu }=\frac{\beta _{2}^{(1)}}{\chi _{2}^{(1)}}\,I^{\mu
}\;,\;\;\;\;X_{1}^{\mu \nu }=\frac{\beta _{1}^{(2)}}{\chi _{1}^{(2)}}%
\,\sigma ^{\mu \nu }\;.  \label{crucialstep}
\end{equation}%
These equations are then used to express Eqs.~(\ref{motion}) in terms of
only $n^{\mu }$ and $\pi ^{\mu \nu }$ (up to some order in \textrm{Kn} and 
\textrm{R}$^{-1}$), leading to the following set of dynamical equations (in
the massless limit) \cite{DNMR},%
\begin{align}
\tau _{n}\dot{n}^{\left\langle \mu \right\rangle }+n^{\mu }& =\kappa
I^{\mu }+\mathcal{J}^{\mu }+\mathcal{K}^{\mu }+\mathcal{R}^{\mu }\text{ }%
,  \notag \\
\tau _{\pi }\dot{\pi}^{\left\langle \mu \nu \right\rangle }+\pi ^{\mu \nu }&
=2\eta \sigma ^{\mu \nu }+\mathcal{J}^{\mu \nu }+\mathcal{K}^{\mu \nu }+%
\mathcal{R}^{\mu \nu }\;.  \label{Final}
\end{align}%
In the above equations of motion, all nonlinear terms and couplings to other
currents were collected in the tensors $\mathcal{J}^{\mu }$, $\mathcal{K}%
^{\mu }$, $\mathcal{R}^{\mu }$, $\mathcal{J}^{\mu \nu }$, $\mathcal{K}^{\mu
\nu }$, and $\mathcal{R}^{\mu \nu }$. The tensors $\mathcal{J}^{\mu }$ and $%
\mathcal{J}^{\mu \nu }$ contain all terms of first order in the product of
Knudsen and inverse Reynolds numbers,%
\begin{align}
\mathcal{J}^{\mu }& =-n_{\nu }\omega ^{\nu \mu }-\delta _{nn}n^{\mu }\theta
+\ell _{n\pi }\Delta ^{\mu \nu }\nabla _{\lambda }\pi _{\nu }^{\lambda
}-\tau _{n\pi }\pi ^{\mu \nu }F_{\nu }-\lambda _{nn}n_{\nu }\sigma ^{\mu \nu
}-\lambda _{n\pi }\pi ^{\mu \nu }I_{\nu }\;,  \notag \\
\mathcal{J}^{\mu \nu }& =2\pi _{\lambda }^{\left\langle \mu \right. }\omega
^{\left. \nu \right\rangle \lambda }-\delta _{\pi \pi }\pi ^{\mu \nu }\theta
-\tau _{\pi \pi }\pi ^{\lambda \left\langle \mu \right. }\sigma _{\lambda
}^{\left. \nu \right\rangle }-\tau _{\pi n}n^{\left\langle \mu \right.
}F^{\left. \nu \right\rangle }+\ell _{\pi n}\nabla ^{\left\langle \mu
\right. }n^{\left. \nu \right\rangle }+\lambda _{\pi n}n^{\left\langle \mu
\right. }I^{\left. \nu \right\rangle }\;.  \label{14_moment_terms}
\end{align}%
where we defined $\theta =\nabla _{\mu }u^{\mu }$, $F^{\mu }=\nabla ^{\mu
}P_{0}$, and $\omega ^{\mu \nu }=\left( \nabla ^{\mu }u^{\nu }-\nabla ^{\nu
}u^{\mu }\right) /2$. The tensors $\mathcal{K}^{\mu }$ and $\mathcal{K}^{\mu
\nu }$ contain all terms of second order in Knudsen number,%
\begin{align}
\mathcal{K}^{\mu }& =\bar{\kappa} _{1}\sigma ^{\mu \nu }I_{\nu }+\bar{\kappa} _{2}\sigma
^{\mu \nu }F_{\nu }+\bar{\kappa} _{3}I^{\mu }\theta +\bar{\kappa} _{4}F^{\mu }\theta
+\bar{\kappa} _{5}\omega ^{\mu \nu }I_{\nu }+\bar{\kappa} _{6}\Delta _{\lambda }^{\mu
}\partial _{\nu }\sigma ^{\lambda \nu }+\bar{\kappa} _{7}\nabla ^{\mu }\theta , 
\notag \\
\mathcal{K}^{\mu \nu }& =\bar{\eta} _{1}\omega _{\lambda }^{\left. {}\right.
\left\langle \mu \right. }\omega ^{\left. \nu \right\rangle \lambda }+\bar{\eta}
_{2}\theta \sigma ^{\mu \nu }+\bar{\eta} _{3}\sigma ^{\lambda \left\langle \mu
\right. }\sigma _{\lambda }^{\left. \nu \right\rangle }+\bar{\eta} _{4}\sigma
_{\lambda }^{\left\langle \mu \right. }\omega ^{\left. \nu \right\rangle
\lambda }+\bar{\eta} _{5}I^{\left\langle \mu \right. }I^{\left. \nu \right\rangle
}+\bar{\eta} _{6}F^{\left\langle \mu \right. }F^{\left. \nu \right\rangle }+\bar{\eta}
_{7}I^{\left\langle \mu \right. }F^{\left. \nu \right\rangle }+\bar{\eta}
_{8}\nabla ^{\left\langle \mu \right. }I^{\left. \nu \right\rangle }+\bar{\eta}
_{9}\nabla ^{\left\langle \mu \right. }F^{\left. \nu \right\rangle }.
\end{align}%
The tensors $\mathcal{R}^{\mu }$ and $\mathcal{R}^{\mu \nu }$ contain all
terms of second order in inverse Reynolds number,%
\begin{equation}
\mathcal{R}^{\mu }=\varphi _{4}n_{\nu }\pi ^{\mu \nu },\text{ \ \ }\mathcal{R%
}^{\mu \nu }=\varphi _{7}\pi ^{\lambda \left\langle \mu \right. }\pi
_{\lambda }^{\left. \nu \right\rangle }+\varphi _{8}n^{\left\langle \mu
\right. }n^{\left. \nu \right\rangle }.
\end{equation}%
These terms only appear when we consider nonlinear contributions in $\delta
f_{\mathbf{k}}$ in the expansion of the single-particle distribution
functions appearing in the collision term. Here, we will not consider such
nonlinear contributions. In Eqs.\ (\ref{Final}), terms of order $\mathcal{O}(%
\mathrm{Kn}^{3})$, $\mathcal{O}(\mathrm{R}^{-3})$, $\mathcal{O}(\mathrm{Kn}%
^{2}\mathrm{R}^{-1})$ and $\mathcal{O}(\mathrm{Kn}\,\mathrm{R}^{-2})$ were
omitted.

Up to order $\mathcal{O}(\mathrm{R}^{-1})$, $\mathcal{O}(\mathrm{Kn})$ and $%
\mathcal{O}(\mathrm{R}^{-1}\mathrm{Kn})$, Eqs.\ (\ref{Final}) have the same
structure as those derived within IS theory [if all terms are kept in the
derivation \cite{Betz}], although with different transport coefficients \cite%
{DNMR}: in principle, in RTRFD the transport coefficients carry information
from \emph{all\/} moments of the distribution function. In practice, the
precision required for the values of the transport coefficients sets a limit
on the number of moments that actually need to be taken into account. We
found that, for the purposes of this work, 37 moments turn out to be
sufficient.

We now note that there is a crucial problem with Eqs.\ (\ref{Final}): the
terms of higher order in Knudsen number, e.g.\ $\nabla ^{\left\langle \mu
\right. }I^{\left. \nu \right\rangle }$, $\nabla ^{\left\langle \mu \right.
}F^{\left. \nu \right\rangle }$, $\Delta _{\alpha }^{\mu }\partial _{\nu
}\sigma ^{\alpha \nu }$ and $\nabla ^{\mu }\theta $, have second-order
spatial derivatives and thus render these equations parabolic, i.e.,
acausal. Note that this happens despite the introduction of the slowest
microscopic time scales as relaxation times for the particle diffusion
current and the shear-stress tensor. Therefore, in order to solve Eqs.\ (\ref%
{Final}) in a relativistic setting, we have to neglect (at least some of)
the terms (in) $\mathcal{K}^\mu,\, \mathcal{K}^{\mu \nu}$.

However, there are situations where these terms are important and should not
be simply neglected. For instance, gradients of shear stress (the term $\sim
\bar{\kappa}_{6}\Delta _{\lambda }^{\mu }\partial _{\nu }\sigma ^{\lambda \nu }
\subset \mathcal{K}^\mu$) and of the shear-stress tensor (the term $\sim
\ell _{n\pi }\Delta ^{\mu \nu }\nabla _{\lambda }\pi _{\nu }^{\lambda}
\subset \mathcal{J}^\mu$) could be of the same order, resulting in source
terms of similar magnitude in the equation for the particle diffusion
current. Neglecting the former can spoil the agreement with numerical
solutions of the Boltzmann equation, especially in the case where the
Navier-Stokes term $\sim I^\mu$ is small compared to these second-order
terms. Similar arguments apply to a situation where the gradients of $I^\mu$
and of the particle diffusion current are of the same order of magnitude and
the Navier-Stokes term $\sim \sigma^{\mu \nu}$ is small. We shall study
these situations in detail in Sec.\ \ref{Comp}.

We remark that this is also prone to happen in cases where the Navier-Stokes
terms are of different order of magnitude. This would require the
introduction of different Knudsen numbers for the various dissipative
currents and the previous power-counting scheme has to be adapted to this
situation. In this work, we do not pursue this avenue further from a formal
point of view. Rather, we discuss in the next section how to include
second-order terms in Knudsen number in a way that preserves hyperbolicity.

\section{Causal transient Fluid dynamics up to second order in Knudsen number%
}

\label{Hyper}

The parabolic and, thus, acausal nature of the equations of motion (\ref%
{Final}) can be understood as follows. The main assumption of RTRFD is to
approximate the quickly varying eigenmodes of the Boltzmann equation by
their asymptotic (i.e., Navier-Stokes) values. The approximation of the
quickly varying eigenmodes with $r\geq 1$ happened in Eq.\ (\ref{asymptotic}),
while the substitution of the eigenmodes with $r=1$ (for $\ell=2$) and
$r=2$ (for $\ell =1$) by their Navier-Stokes
values occurred in Eq.\ (\ref{crucialstep}). However, in this step it was
implicitly assumed that these eigenmodes relax instantaneously to their
corresponding Navier-Stokes values, consequently leading to acausal behavior
and, ultimately, to the parabolic terms in Eqs.\ (\ref{Final}).

In order to obtain hyperbolic equations of motion which do not simply
neglect terms of order $\mathcal{O}(\mathrm{Kn}^{2})$, it is necessary to
refrain from the substitution (\ref{crucialstep}). This can be simply done
by keeping $X_{2}^{\mu }$ and $X_{1}^{\mu \nu }$ in Eqs.\ (\ref{noname}) as
independent dynamical variables instead of replacing them by their
Navier-Stokes values. However, we do not solve differential equations of the
type (\ref{help formulas}) for these variables. Rather, for a given $r$ we
solve the first Eq.\ (\ref{noname}) for $X_{2}^{\mu }$ and, for a given $s$
(not necessarily equal to $r$) the second Eq.\ (\ref{noname}) for $%
X_{1}^{\mu \nu }$. Thus, we can replace these variables by a set of
irreducible moments $\rho _{r}^{\mu }$, $\rho _{s}^{\mu \nu }$. Without loss
of generality, we choose $r=2$ and $s=1$, i.e., the irreducible moments $%
\rho _{2}^{\mu }$ and $\rho _{1}^{\mu \nu }$ as independent dynamical
variables. In this way, we obtain 
\begin{eqnarray}
\rho _{r}^{\mu } &=&\text{ }\lambda _{r0}^{\left( 1\right) }n^{\mu }+\lambda
_{r2}^{\left( 1\right) }\rho _{2}^{\mu }+\mathcal{O}(\mathrm{Kn}^{2},\mathrm{%
Kn}\text{ }\mathrm{R}^{-1},\mathrm{R}^{-2})\text{ },  \notag \\
\rho _{r}^{\mu \nu } &=&\lambda _{r0}^{\left( 2\right) }\pi ^{\mu \nu
}+\lambda _{r1}^{\left( 2\right) }\rho _{1}^{\mu \nu }+\mathcal{O}(\mathrm{Kn%
}^{2},\mathrm{Kn}\text{ }\mathrm{R}^{-1},\mathrm{R}^{-2})\text{ },
\label{dontcare}
\end{eqnarray}%
where we defined 
\begin{equation}
\lambda _{r0}^{\left( 1\right) }=\frac{\Omega _{20}^{\left( 1\right) }\kappa
_{r}-\Omega _{r0}^{\left( 1\right) }\kappa _{2}}{\Omega _{20}^{\left(
1\right) }\kappa _{0}-\kappa _{2}},\text{ \ }\lambda _{r2}^{\left(
1\right) }=\frac{\kappa _{0}\Omega _{r0}^{\left( 1\right) }-\kappa _{r}%
}{\kappa _{0}\Omega _{20}^{\left( 1\right) }-\kappa _{2}},\text{ }%
\lambda _{r0}^{\left( 2\right) }=\frac{\Omega _{10}^{\left( 2\right) }\eta
_{r}-\Omega _{r0}^{\left( 2\right) }\eta _{1}}{\Omega _{10}^{\left( 2\right)
}\eta _{0}-\eta _{1}},\text{ \ }\lambda _{r1}^{\left( 2\right) }=\frac{%
\Omega _{r0}^{\left( 2\right) }\eta _{0}-\eta _{r}}{\Omega _{10}^{\left(
2\right) }\eta _{0}-\eta _{1}}\text{ .}
\end{equation}%
These relations hold for any $r$ and, therefore, up to order $\mathcal{O}(%
\mathrm{Kn}^{2},\mathrm{Kn}$ $\mathrm{R}^{-1},\mathrm{R}^{-2})$ it is
possible to approximate all irreducible moments solely in terms of $n^{\mu }$%
, $\rho _{2}^{\mu }$, $\pi ^{\mu \nu }$, and $\rho _{1}^{\mu \nu }$.

Note, however, that the relations (\ref{dontcare}) are only valid for the
irreducible moments $\rho _{r}^{\mu_1 \cdots \mu_\ell }$ with positive $r$.
Nevertheless, similar relations can also be obtained for the irreducible
moments with negative $r$. This can be done by substituting Eq.\ (\ref%
{fexpansion2}) into Eq.\ (\ref{rho}), thereby expressing the irreducible moments
with negative $r$ in terms of those with positive $r$, 
\begin{equation}
\rho _{-r}^{\nu _{1}\cdots \nu _{\ell }}=\sum_{n=0}^{N_{\ell }}\mathcal{F}%
_{rn}^{\left( \ell \right) }\rho _{n}^{\nu _{1}\cdots \nu _{\ell }},
\label{TTT}
\end{equation}%
where we defined the following thermodynamic integral%
\begin{equation}
\mathcal{F}_{rn}^{\left( \ell \right) }=\frac{\ell !}{\left( 2\ell +1\right)
!!}\int dK\text{ }f_{0\mathbf{k}}\tilde{f}_{0\mathbf{k}}E_{\mathbf{k}}^{-r}%
\mathcal{H}_{\mathbf{k}n}^{\left( \ell \right) }\left( \Delta ^{\alpha \beta
}k_{\alpha }k_{\beta }\right) ^{\ell }.
\end{equation}%
Then, we approximate $\rho _{-r}^{\mu }$ (in terms of $n^{\mu }$ and $\rho
_{2}^{\mu }$) and $\rho _{-r}^{\mu \nu }$ (in terms of $\pi ^{\mu \nu }$ and 
$\rho_1 ^{\mu \nu }$) by substituting Eqs.\ (\ref{dontcare}) into Eq.\ (\ref{TTT}%
), leading to%
\begin{eqnarray}
\rho _{-r}^{\mu } &=&\left[ \sum_{n=0,\neq 1}^{N_{1}}\mathcal{F}%
_{rn}^{\left( 1\right) }\lambda _{n0}^{\left( 1\right) }\right] n^{\mu }+%
\left[ \sum_{n=0,\neq 1}^{N_{1}}\mathcal{F}_{rn}^{\left( 1\right) }\lambda
_{n2}^{\left( 1\right) }\right] \rho _{2}^{\mu }+\mathcal{O}(\mathrm{Kn}^{2},%
\mathrm{Kn}\text{ }\mathrm{R}_{i}^{-1},\mathrm{R}_{i}^{-2}),  \notag \\
\rho _{-r}^{\mu \nu } &=&\left[ \sum_{n=0}^{N_{2}}\mathcal{F}_{rn}^{\left(
2\right) }\lambda _{n0}^{\left( 2\right) }\right] \pi ^{\mu \nu }+\left[
\sum_{n=0}^{N_{2}}\mathcal{F}_{rn}^{\left( 2\right) }\lambda _{n1}^{\left(
2\right) }\right] \rho _{1}^{\mu \nu }+\mathcal{O}(\mathrm{Kn}^{2},\mathrm{Kn%
}\text{ }\mathrm{R}_{i}^{-1},\mathrm{R}_{i}^{-2}).  \label{OMG23}
\end{eqnarray}

Using these relations, Eqs.\ (\ref{motion}) can then be closed in terms of $%
n^{\mu }$, $\pi ^{\mu \nu }$, $\rho _{2}^{\mu }$, and $\rho _{1}^{\mu \nu }$.
In order to derive the equations of motion, we first multiply Eqs.\ (\ref%
{motion}) by $\tau _{nr}^{\left( \ell \right) }$ and sum over $r$. Next, we
use Eqs.\ (\ref{dontcare}) and (\ref{OMG23}) to replace all irreducible
moments $\rho _{i}^{\mu }$ and $\rho _{i}^{\mu \nu }$ appearing in the
equations by $n^{\mu }$, $\rho _{2}^{\mu }$, $\pi ^{\mu \nu }$, and
$\rho_1^{\mu \nu }$. The scalar irreducible moments and those with rank larger than
two are replaced by zero. Additionally, all covariant time derivatives of $%
\alpha _{0}$, $\beta _{0}$, and $u^{\mu }$ are replaced by spatial gradients
of fluid-dynamical variables using the conservation laws, $\partial _{\mu
}N^{\mu }=0,\ \partial _{\mu }T^{\mu \nu }=0$. Then we obtain,%
\begin{align}
\hat{\tau}_{n}\Delta _{\alpha }^{\mu }\frac{d\vec{n}^{\alpha }}{d\tau }+\vec{%
n}^{\mu }& =\vec{\kappa}I^{\mu }-\hat{\tau}_{n}\vec{n}_{\nu }\omega ^{\nu
\mu }-\hat{\delta}_{nn}\vec{n}^{\mu }\theta +\hat{\ell}_{n\pi }\Delta ^{\mu
\nu }\partial _{\lambda }\vec{\pi}_{\nu }^{\lambda }-\hat{\tau}_{n\pi }\vec{%
\pi}^{\mu \nu }F_{\nu }-\hat{\lambda}_{nn}\vec{n}_{\nu }\sigma ^{\mu \nu }-%
\hat{\lambda}_{n\pi }\vec{\pi}^{\mu \nu }I_{\nu }\;,  \notag \\
\hat{\tau}_{\pi }\Delta _{\alpha \beta }^{\mu \nu }\frac{d\vec{\pi}^{\alpha
\beta }}{d\tau }+\vec{\pi}^{\mu \nu }& =2\vec{\eta}\sigma ^{\mu \nu }+2\hat{%
\tau}_{\pi }\vec{\pi}_{\lambda }^{\left\langle \mu \right. }\omega ^{\left.
\nu \right\rangle \lambda }-\hat{\delta}_{\pi \pi }\vec{\pi}^{\mu \nu
}\theta -\hat{\tau}_{\pi \pi }\vec{\pi}^{\lambda \left\langle \mu \right.
}\sigma _{\lambda }^{\left. \nu \right\rangle }\;-\hat{\tau}_{\pi n}\vec{n}%
^{\left\langle \mu \right. }F^{\left. \nu \right\rangle }+\hat{\ell}_{\pi
n}\nabla ^{\left\langle \mu \right. }\vec{n}^{\left. \nu \right\rangle }+%
\hat{\lambda}_{\pi n}\vec{n}^{\left\langle \mu \right. }I^{\left. \nu
\right\rangle }\text{ },  \label{motion2}
\end{align}%
where we defined the vectors,%
\begin{equation}
\vec{n}^{\mu }=\left( 
\begin{array}{c}
n^{\mu } \\ 
\rho _{2}^{\mu }%
\end{array}%
\right) ,\text{ \ }\vec{\pi}^{\mu \nu }=\left( 
\begin{array}{c}
\pi ^{\mu \nu } \\ 
\rho _{1}^{\mu \nu }%
\end{array}%
\right) .
\end{equation}%
In this approximation, RTRFD becomes a theory with 21 moments as dynamical
variables (had we included the scalar moments, there would have been 23
moments) while, in the previous approximation, i.e., Eqs.\ (\ref{Final}),
there were only 13 moments (including bulk viscous pressure, it would have
been 14 moments). Above, $\hat{\tau}_{n}$, $\hat{\tau}_{\pi }$, $\hat{\ell}%
_{n\pi }$, $\hat{\ell}_{\pi n}$, $\hat{\delta}_{nn}$, $\hat{\delta}_{\pi \pi
}$, $\hat{\tau}_{n\pi }$, $\hat{\tau}_{\pi \pi }$, $\hat{\tau}_{\pi n}$, $%
\hat{\lambda}_{nn}$, $\hat{\lambda}_{n\pi }$, and $\hat{\lambda}_{\pi n}$
are $2\times 2$ matrices, while $\vec{\kappa}$ and $\vec{\eta}$ are two-component
vectors. The microscopic formulas for these transport coefficients, for a
massless gas of particles, are shown in Appendix \ref{transport coefficients}%
. Here, they are computed for a gas of classical particles with a constant
cross section $\sigma $, choosing $N_{1}=4$ and $N_{2}=3$. The values for
the diffusion and viscosity coefficients,~$\vec{\kappa}$ and $\vec{\eta}$,
and for the relaxation time matrices, $\hat{\tau}_{n}$ and $\hat{\tau}_{\pi }
$, are,%
\begin{gather}
\frac{\vec{\kappa}}{\lambda _{\mathrm{mfp}}n_{0}}=\left( 
\begin{array}{c}
0,1596 \\ 
-2.3616\text{ }T^{2}%
\end{array}%
\right) ,\text{ }\frac{\vec{\eta}}{\lambda _{\mathrm{mfp}}P_{0}}=\left( 
\begin{array}{c}
1.268 \\ 
6.929\text{ }T%
\end{array}%
\right) ,  \notag \\
\frac{\hat{\tau}_{n}}{\lambda _{\mathrm{mfp}}}=\left( 
\begin{array}{cc}
1.295\left. {}\right. \text{ } & -0.053/T^{2} \\ 
5.18\text{ }T^{2}\left. {}\right.  & 2.787%
\end{array}%
\right) ,\text{ }\frac{\hat{\tau}_{\pi }}{\lambda _{\mathrm{mfp}}}=\left( 
\begin{array}{cc}
0.912\left. {}\right.  & 0.136/T \\ 
-3.647\text{ }T\left. {}\right.  & 2.456\text{ }%
\end{array}%
\right) .
\end{gather}%
The transport coefficients of the nonlinear terms in the equation of motion
for $\vec{n}^{\mu }$ are%
\begin{gather}
\frac{\hat{\delta}_{nn}}{\lambda _{\mathrm{mfp}}}=\left( 
\begin{array}{cc}
1.295\left. {}\right.  & -0.0883/T^{2} \\ 
5.18\text{ }T^{2}\left. {}\right.  & 4.645\left. {}\right. 
\end{array}%
\right) \text{ },\text{\ }\frac{\hat{\lambda}_{nn}}{\lambda _{\mathrm{mfp}}}%
=\left( 
\begin{array}{cc}
0.524\left. {}\right.  & -0.0341/T^{2} \\ 
2.096\text{ }T^{2}\left. {}\right.  & 2.863%
\end{array}%
\right) ,\text{ }\frac{\hat{\lambda}_{n\pi }}{\lambda _{\mathrm{mfp}}}%
=\left( 
\begin{array}{cc}
0.1677/T\left. {}\right.  & -0.0288/T^{2} \\ 
0.6708\text{ }T\left. {}\right.  & -0.1147\left. {}\right. 
\end{array}%
\right) ,  \notag \\
\frac{\hat{\tau}_{n\pi }}{\lambda _{\mathrm{mfp}}}=\frac{1}{4P_{0}}\left( 
\begin{array}{cc}
0\left. {}\right. \text{ } & 0.0973/T^{2} \\ 
0\left. {}\right.  & -2.6106\left. {}\right. 
\end{array}%
\right) ,\text{ }\frac{\hat{\ell}_{n\pi }}{\lambda _{\mathrm{mfp}}}=\left( 
\begin{array}{cc}
-0.4723/T\left. {}\right.  & 0.0973/T^{2} \\ 
13.111\text{ }T\left. {}\right.  & -2.611\left. {}\right. 
\end{array}%
\right) ,
\end{gather}%
while those in the equation of motion for $\vec{\pi}^{\mu \nu }$ are 
\begin{gather}
\frac{\hat{\delta}_{\pi \pi }}{\lambda _{\mathrm{mfp}}}=-\frac{4}{3}\left( 
\begin{array}{cc}
0.912\left. {}\right.  & 0.17/T \\ 
-3.647\text{ }T\left. {}\right.  & 3.0698\text{ }%
\end{array}%
\right) ,\text{ }\frac{\hat{\tau}_{\pi \pi }}{\lambda _{\mathrm{mfp}}}%
=\left( 
\begin{array}{cc}
1.5688\left. {}\right.  & 0.2261/T \\ 
-6.2751\text{ }T\left. {}\right.  & 5.0956%
\end{array}%
\right) ,\text{ }\frac{\hat{\tau}_{\pi n}}{\lambda _{\mathrm{mfp}}}=\frac{1}{%
P_{0}}\left( 
\begin{array}{cc}
0.2228\text{ }T\left. {}\right. \text{ } & 0.0714/T \\ 
0.8913\text{ }T^{2}\left. {}\right.  & 1.5144%
\end{array}%
\right) ,  \notag \\
\frac{\hat{\ell}_{\pi n}}{\lambda _{\mathrm{mfp}}}=\left( 
\begin{array}{cc}
0.2228\text{ }T\text{ }\left. {}\right.  & 0.0476/T \\ 
0.8913\text{ }T^{2}\left. {}\right.  & 1.0096%
\end{array}%
\right) ,\text{ }\frac{\hat{\lambda}_{\pi n}}{\lambda _{\mathrm{mfp}}}%
=\left( 
\begin{array}{cc}
0.1186\text{ }T\text{ } & 0.0084/T \\ 
-0.4744\text{ }T^{2} & -0.0338%
\end{array}%
\right) \text{ }.
\end{gather}

Note that these equations of motion are hyperbolic and include terms of $%
\mathcal{O}(\mathrm{Kn}^{2})$. We remark that, while this approach increases
the domain of validity of the equations of motion without making them
parabolic, it does not solve the intrinsic problem of the coarse-graining
procedure: if one attempts to go to an even higher order in Knudsen number,
the equations become once more parabolic. This can be solved in a similar
fashion, by again increasing the number of irreducible moments describing
the system.

\section{Comparison with the Boltzmann equation}

\label{Comp}

In order to test the validity of RTRFD, we compare it to the solution of the
Boltzmann equation, similarly as in Ref.\ \cite{Bouras}. The numerical
method of choice for solving the Boltzmann equation is the Boltzmann
approach for multiparton scattering (BAMPS) \cite{BAMPS}, while the
macroscopic field equations are solved using the viscous SHarp And Smooth
Transport Algorithm (vSHASTA) \cite{Molnar:2009tx}. Both fluid dynamics and
the Boltzmann equation are solved in Cartesian coordinates with a flat
space-time metric $g^{\mu \nu }=\mathrm{diag}(1,-1,-1,-1)$.

\begin{figure}[th]
\hspace{-0.0cm} \includegraphics[width=12.0cm]{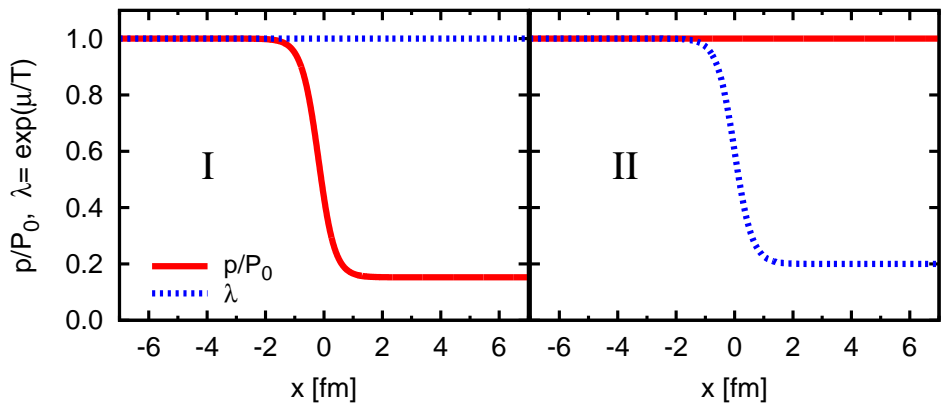} 
\caption{{\protect\small (Color online) Initial conditions for cases I and
II. }}
\label{fig:inicond}
\end{figure}

We consider a (3+1)--dimensional system, but assume that the matter is
homogeneous in the $y-z$ plane, allowing for an inhomogeneous matter
distribution only in the longitudinal $x$--direction. This effectively leads
to a (1+1)--dimensional problem. We consider two different initial
conditions. In case I, the system is initialized with a homogeneous
fugacity, $\lambda_0 =e^{\alpha_{0}}\equiv 1$, but with an inhomogeneous
pressure profile in the longitudinal direction. In practice we
smoothly connect two temperature states $T_{(-\infty)} = 0.4 \, \rm{GeV}$
and $T_{(+\infty)}= 0.25 \, \rm{GeV}$ via the Woods-Saxon parametrization
with thickness parameter $D = 0.3$ fm. In case II, the 
pressure $P_0 = g {T_{(-\infty)}}^4 / \pi^2$
is homogeneous and the fugacity distribution is also given by a Woods-Saxon profile
with $D = 0.3$ fm, interpolating between $\lambda_{(-\infty)} = 1$ and
$\lambda_{(+\infty)}= 0.2$. For the degeneracy factor we use $g = 16$. In both
cases, matter is initialized in local thermodynamical equilibrium, i.e.,
with all dissipative currents (and eigenmodes of the Boltzmann equation) set
to zero, and at rest, i.e., with a vanishing collective velocity $u^{\mu }=0$.
These initial conditions are shown in Fig.\ \ref{fig:inicond}.

%

In both cases, we consider two exemplary values for the cross section $%
\sigma =2$ and $8$ mb, and consider the solutions after the system has
evolved for $6$ fm in time. We compare the solution of the Boltzmann
equation with that of traditional IS theory [including terms omitted in the
original work \cite{IS} but quoted in Ref.\ \cite{Betz}], as well as with
RTRFD at various levels of approximation. Equations (\ref{Final}) contain 13
moments as independent dynamical variables (and 14, if we include bulk
viscous pressure). The calculation of the transport coefficients in these
equations can be done with increasing accuracy, as more irreducible moments
are considered in the expansion (\ref{fexpansion2}). The lowest possible
accuracy is reached if no more than the original 13 (14) irreducible moments
are considered for the calculation of the transport coefficients. At the
next level, we include one more set of irreducible moments of tensor-rank
one and two (and one more scalar moment in the case of non-vanishing bulk
viscous pressure), which leads to a total of 21 (23) irreducible moments. In
this way, the number of irreducible moments entering the transport
coefficients increases by 8 (9) at each successive level of approximation.
For the purpose of this work, we found that going to the third level of
iteration, i.e., considering $13+8\cdot 3=37$ moments ($14+9\cdot 3 = 41$ in
the case of nonvanishing bulk viscous pressure) is sufficient to reach the
desired accuracy in the values of the transport coefficients. In the
following, we shall compare RTRFD with 13 dynamical degrees of freedom and
with the transport coefficients computed with 13 and with 37 moments. We
shall term these variants of RTRFD ``13/13'' and ``13/37''. In addition, we
also solve Eqs.\ (\ref{motion2}). These contain 21 dynamical degrees of
freedom. We compute the corresponding transport coefficients using 37
moments. We shall refer to this variant of RTRFD as ``21/37''.

In the following figures, the numerical solutions of the Boltzmann equation
shall always be displayed by open dots, the results of IS theory by black
dash-dotted lines, the solution of RTRFD ``13/13'' by green dashed lines,
that of RTRFD ``13/37'' by blue dotted lines, and that of RTRFD ``21/37'' by
solid red curves.

In order to verify the different fluid-dynamical theories discussed in
this paper the solutions of the Boltzmann equation must be
calculated to a very high precisison. For this purpose we
performed $5 \cdot 10^{4}$ BAMPS runs and computed the fluid-dynamical
quantities as averages over these runs.

\begin{figure}[th]
\hspace{-0.0cm} \includegraphics[width=12cm]{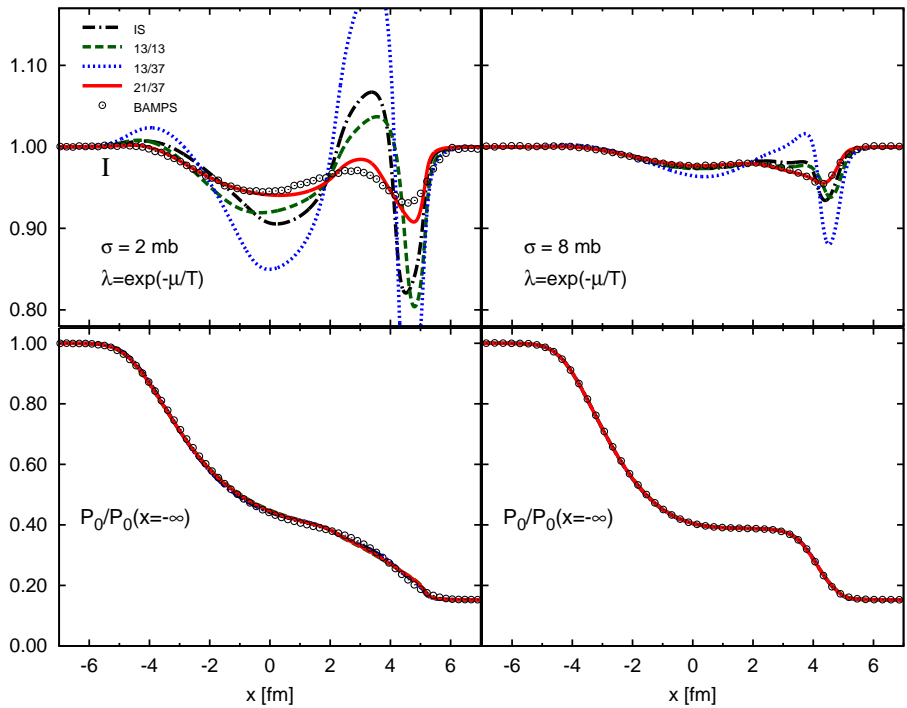} 
\caption{{\protect\small (Color online) Fugacity and thermodynamic pressure
profiles at $t=6$ fm for case I, for $\protect\sigma = 2$ mb (left panels)
and $\protect\sigma = 8$ mb (right panels) .}}
\label{fig:caseI}
\end{figure}

\begin{figure}[th]
\hspace{-0.0cm} \includegraphics[width=12cm]{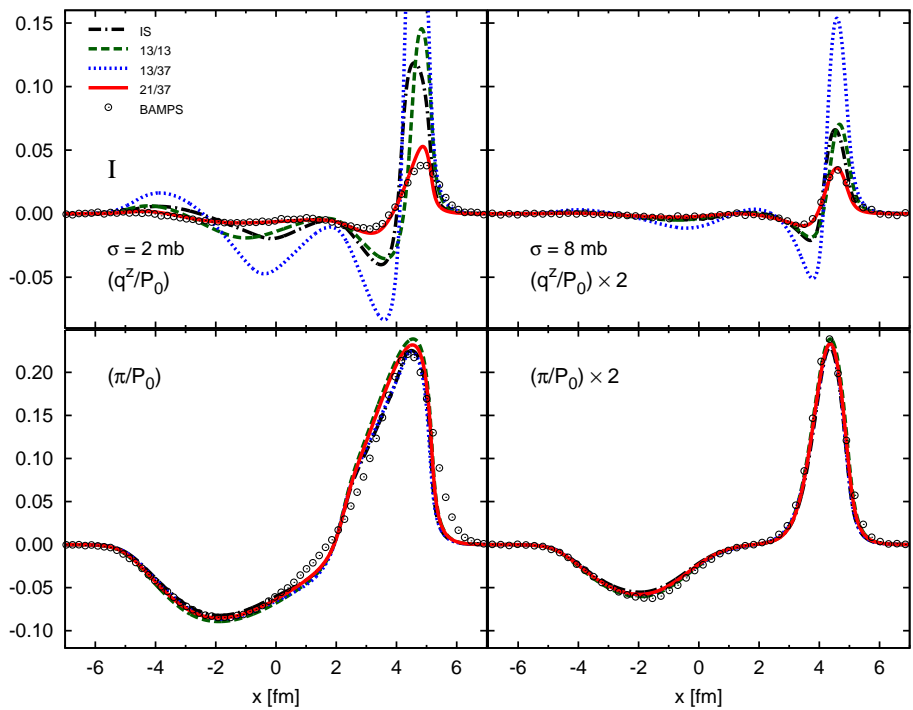} 
\caption{{\protect\small (Color online) Shear-stress tensor and heat flow
profiles at $t=6$ fm for case I, for $\protect\sigma = 2$ mb (left panels)
and $\protect\sigma = 8$ mb (right panels).}}
\label{fig:caseIqp}
\end{figure}

In Fig.\ \ref{fig:caseI} we show the fugacity (top) and thermodynamic
pressure (bottom) and in Fig.\ \ref{fig:caseIqp} the heat flow $q^{\mu
}\equiv -(\varepsilon+P_{0})n^{\mu }/n$ (top) and shear-stress tensor
(bottom) for case I. The Boltzmann equation and the fluid-dynamical theories
were solved for $\sigma =2$ mb (shown in the left panels of each figure) and
for $\sigma =8$ mb (shown in the right panels). For $\sigma =8$ mb, the
thermodynamic pressure and shear-stress tensor computed in all
fluid-dynamical theories are in good agreement with the numerical solutions
of the Boltzmann equation. As we decrease the cross section we expect the
agreement between macroscopic and microscopic theory to become worse. This
explains why, for $\sigma = 2$ mb, the pressure and shear-stress tensor
computed within fluid-dynamical theories deviate more strongly from those
computed via the microscopic theory. Nevertheless, compared to the fugacity
and heat-flow profiles, the agreement is not too bad, even for the smaller
value of the cross section.

The initial pressure gradient in case I drives, via conservation of
momentum, the creation of velocity gradients. On the other hand, the
gradient of fugacity is initially zero and turns out to remain small
throughout the evolution. In this situation, higher-order terms involving
gradients of velocity and of the shear-stress tensor, e.g.\ $\bar{\kappa}
_{6}\Delta _{\lambda }^{\mu }\partial _{\nu }\sigma ^{\lambda \nu }\subset 
\mathcal{K}^{\mu }$ and $\ell _{n\pi }\Delta ^{\mu \nu }\nabla _{\lambda
}\pi _{\nu }^{\lambda }\subset \mathcal{J}^{\mu }$ in the particle diffusion
equation (\ref{Final}), become of the same order as the respective
(first-order) Navier-Stokes term $\kappa I^{\mu }$. Therefore, if terms of
this type are not properly taken into account, we expect large deviations
from the solution of the Boltzmann equation. This can be seen in Figs.\ \ref%
{fig:caseI} and \ref{fig:caseIqp} when comparing IS theory, RTRFD
\textquotedblleft 13/13\textquotedblright , as well as RTRFD
\textquotedblleft 13/37\textquotedblright\ with the Boltzmann result. In all
of these variants, the parabolic term $\sim \bar{\kappa} _{6}$ is either absent
(IS theory and RTRFD \textquotedblleft 13/13\textquotedblright ) or has to
be dismissed (RTRFD \textquotedblleft 13/37\textquotedblright ) for reasons
of causality. In addition, IS theory and RTRFD \textquotedblleft
13/13\textquotedblright\ do not have the correct value for $\ell _{n\pi }$,
because we did not include a sufficiently large number of irreducible
moments in its computation. Although RTRFD \textquotedblleft
13/37\textquotedblright\ features (within the desired accuracy) the correct
value for this transport coefficient (as well as for $\bar{\kappa} _{6}$), it does
even worse in describing the fugacity and heat-flow profiles than the
previous two theories. This is because the term $\sim \bar{\kappa} _{6}$ could not
be taken into account for reasons of causality, although it is of the same
order of magnitude as the term $\sim \ell _{n\pi }$. These problems of
fluid-dynamical theories with only 13 dynamical variables are resolved by
RTRFD \textquotedblleft 21/37\textquotedblright\ which is the only
fluid-dynamical theory considered here that contains \textit{all}
contributions of second-order in the Knudsen number in a hyperbolic fashion.

\begin{figure}[th]
\hspace{-0.0cm} \includegraphics[width=12cm]{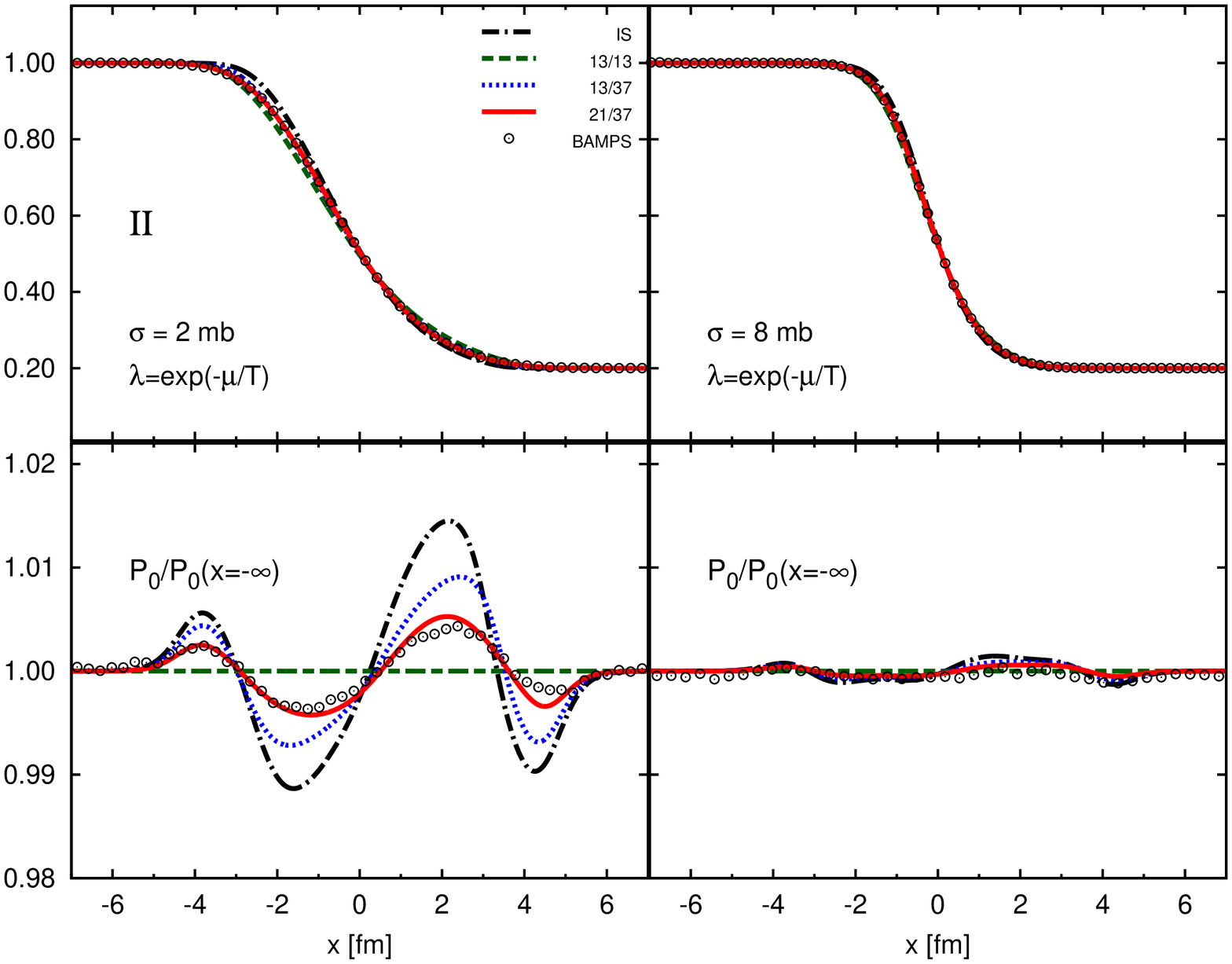} 
\caption{{\protect\small (Color online) Shear-stress tensor and heat flow at 
$t=6$ fm for case II, for $\protect\sigma =2$ mb (left panels) and $\protect%
\sigma =8$ mb (right panels).}}
\label{fig:caseII}
\end{figure}

\begin{figure}[th]
\hspace{-0.0cm} \includegraphics[width=12cm]{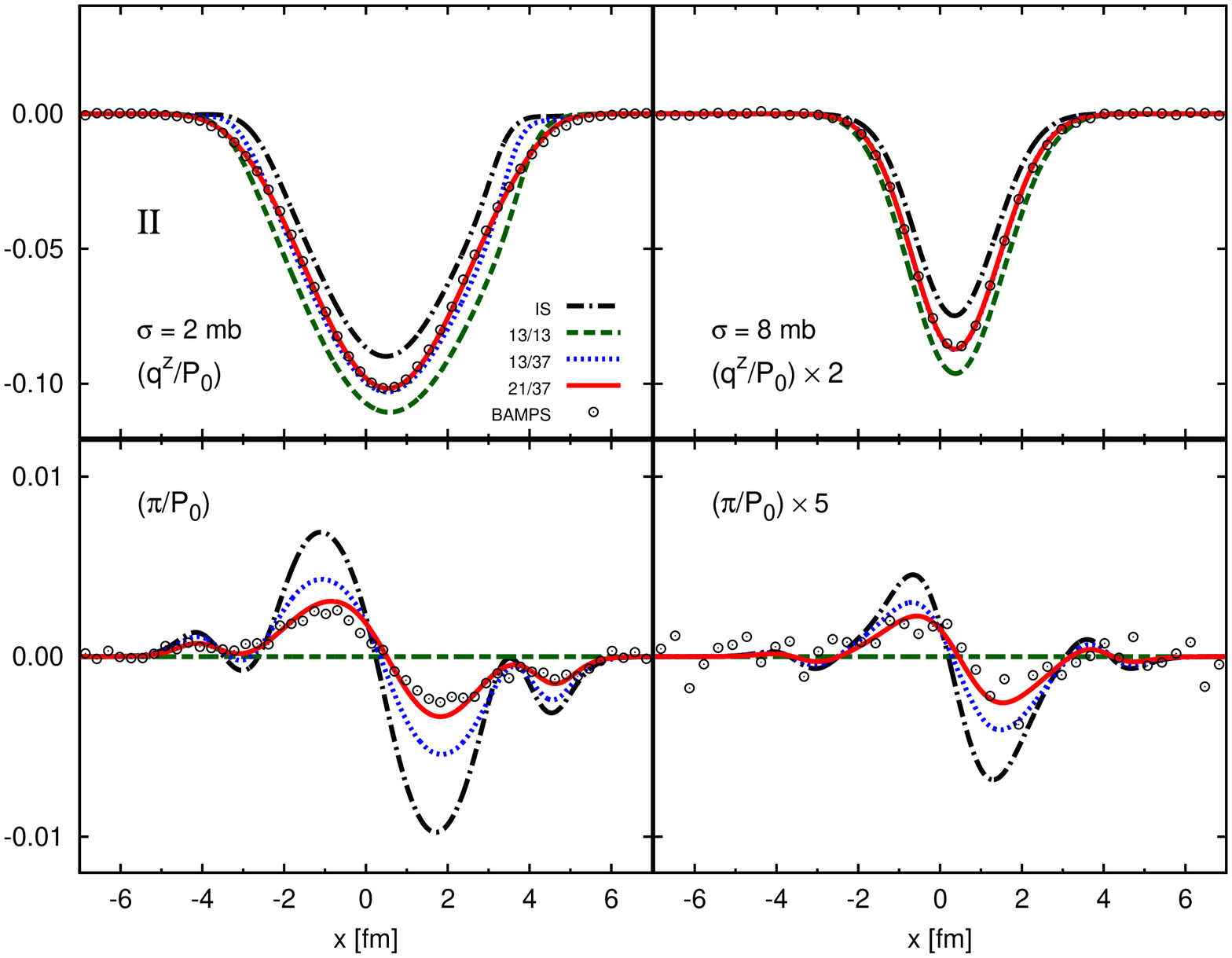} 
\caption{{\protect\small (Color online) Shear-stress tensor and heat flow
profiles at $t=6$ fm for case II, for $\protect\sigma =2$ mb (left panels)
and $\protect\sigma =8$ mb (right panels). }}
\label{fig:caseIIqp}
\end{figure}

In Fig.\ \ref{fig:caseII} we show the fugacity (top) and thermodynamic
pressure (bottom) and in Fig.\ \ref{fig:caseIIqp} the heat-flow (top) and
shear-stress tensor profiles (bottom) for case II. As before, the Boltzmann
equation and the fluid-dynamical theories considered were solved for $\sigma
=2$ mb (shown in the left panels) and for $\sigma =8$ mb (shown in the right
panels). Again, we expect, and see, better agreement between fluid dynamics
and the Boltzmann equation for the larger value of the cross section. While
the fugacity profiles are in good agreement with the solution of the
Boltzmann equation for all fluid-dynamical theories and both values of the
cross section, the heat flow is not well described in IS theory and in RTRFD
\textquotedblleft 13/13\textquotedblright : IS theory predicts values for
the heat flow which are smaller in magnitude than the Boltzmann equation,
while RTRFD \textquotedblleft 13/13\textquotedblright\ predicts larger
values, even for $\sigma =8$ mb. On the other hand, both RTRFD
\textquotedblleft 13/37\textquotedblright\ and RTRFD \textquotedblleft
21/37\textquotedblright\ describe the heat flow very well or even perfectly,
respectively, for both values of the cross section. The reason is that the
diffusion coefficient $\kappa $ has the correct value in these theories
(while it deviates by $\sim 30\%$ in both IS theory and RTRFD
\textquotedblleft 13/13\textquotedblright ).

Since, in case II, the initial pressure gradient is zero and turns out to
remain small throughout the evolution, the velocity gradients remain small
as well. In this situation, it is important to include higher-order terms
that couple the shear-stress tensor to heat flow. This is the reason why the
solutions of IS theory and RTRFD ``13/13'' (where these higher-order terms
vanish in the massless limit) are not in good agreement with that of the
Boltzmann equation for the thermodynamic pressure and the shear-stress
tensor, for both values of the cross section. On the other hand, RTRFD
``13/37'' does a better job in matching the Boltzmann equation. It is not
perfect, because the higher-order terms $\sim \bar{\eta}_5$ and $\sim \bar{\eta}_8$ were
dropped. The best agreement is, again, found within RTRFD ``21/37'' where
all second-order terms in the Knudsen number are taken into account.

Note that, in Fig.\ \ref{fig:caseIIqp} the BAMPS results for 
the shear-stress tensor are strongly fluctuating.
This happens because, in this special case, the values of the shear-stress
tensor are of the same magnitude as the statistical fluctuations in BAMPS.
In order to reduce the statistical fluctuations and
to achieve a better resolution, a significantly larger amount of
runs would be required.

\section{Summary and Conclusion}

\label{Concl}

In this paper, we compared the equations of motion of RTRFD at
various levels of approximation with numerical solutions of the Boltzmann
equation for two different types of initial conditons (labeled case I and
II). Also, we showed how to use the RTRFD formalism to derive equations of
motion that are hyperbolic and, at the same time, include terms up to second
order in the Knudsen number. By a careful comparison with numerical
solutions of the microscopic theory, we demonstrated that this formalism is
able to handle problems with strong initial gradients in pressure or
particle number density.

The initial conditions, cases I and II, were chosen in such a way that
considerably different spatial gradient profiles are generated throughout
the fluid-dynamical evolution. In case I, the pressure gradient is initially
large which gives rise to large velocity gradients in the later stages of
the evolution. This means that, in case I, the shear-stress tensor is mainly
generated by its corresponding Navier-Stokes term, i.e., by gradients of
velocity. On the other hand, the fugacity gradient is initially zero in case
I, and remains relatively small throughout the evolution of the fluid.
Therefore, in case I the heat flow is not mainly created by its
Navier-Stokes term, i.e., by the gradient of fugacity, but by the
coupling term to the shear tensor and shear-stress tensor, i.e., the terms $%
\Delta ^{\mu \nu }\nabla _{\lambda }\pi _{\nu }^{\lambda }$ and $\Delta
^{\mu \nu }\nabla _{\lambda }\sigma _{\nu }^{\lambda }$ in Eqs.\ (\ref{Final}%
). Therefore, in this case the higher-order terms in Knudsen number must be
included and one really needs to solve the hyperbolic equations derived in
this paper to obtain a good agreement. The fact that IS theory is always
deviating from the microscopic theory when it concerns heat flow means that
it does not predict correctly the terms of order one and two in Knudsen
number. On the other hand, in RTRFD \textquotedblleft
13/37\textquotedblright\ and RTRFD \textquotedblleft
21/37\textquotedblright\ all transport coefficients are computed with a
sufficiently large number of irreducible moments. This guarantees that all
terms of the desired order are included and is the reason for the better
agreement of these fluid-dynamical theories with the microscopic theory. The
reason why RTRFD \textquotedblleft 13/37\textquotedblright\ fails in certain
situations is that important terms have to be neglected in order to preserve
hyperbolicity and causality.

In case II, the fugacity gradient is initially large while the pressure
gradient is zero. This means that the heat flow originates mainly from its
Navier-Stokes term while the shear-tress tensor originates mainly from its
coupling to heat flow, i.e., the terms $\nabla ^{<\mu }n^{\nu >}$, $\nabla
^{<\mu }I^{\nu >}$, $I^{<\mu }I^{\nu >}$, and $n^{<\mu }I^{\nu >}$ in Eqs.\ (%
\ref{Final}). The fact that the heat flow calculated from IS theory deviates
from the solution given by the microscopic theory even in this case, is
evidence that the Navier-Stokes term of this theory does not contain the
correct transport coefficient. The coupling of the shear-stress tensor with
the heat flow in IS theory is also not correctly taken into account.

In conclusion, the resummation of irreducible moments for the computation of
the transport coefficients was essential to obtain a good agreement with the
microscopic theory. It provides not only the correct values for the shear
viscosity and heat conduction coefficients, but also for the transport
coefficients that couple the respective dissipative currents. Moreover, in
situations where higher-order terms are important, one has to make sure to
include them in a hyperbolic way, and not simply drop relevant contributions
because they are parabolic. These two factors resolved the previously
observed differences between the solution of IS theory and of the Boltzmann
equation observed in Ref. \cite{Bouras}. 

As expected, and explicitly demonstrated in this paper, the agreement
between solutions of RTRFD and the Boltzmann equation depends on the value
of $\sigma $. For the cases considered in this paper, we obtained a good
agreement for $\sigma =8$ mb, while for $\sigma =2$ mb we started to notice
small deviations. In order to improve the agreement for smaller values of
the cross section, we would have to include more moments of the Boltzmann
equation to describe the state of the system, i.e., such moments would have
to contribute not only to the values of the transport coefficients but also
as independent dynamical variables. We leave this investigation to future
work.

\section{Acknowledgements}

The authors thank H.~Warringa and J.~Noronha for discussions. The work of
H.N.\ was supported by the Extreme Matter Institute (EMMI). E.M is supported
by OTKA/NKTH 81655. This work was supported by Helmholtz International
Center for FAIR within the framework of the LOEWE program launched by the
State of Hesse.

\appendix

\section{Transport coefficients}

\label{transport coefficients}

In this appendix we list all transport coefficients of fluid dynamics
calculated in this paper. The microscopic formulas for the diffusion and
viscosity coefficients,~$\vec{\kappa}$ and $\vec{\eta}$, and for the
relaxation time matrices, $\hat{\tau}_{n}$ and $\hat{\tau}_{\pi }$, are%
\begin{gather}
\vec{\kappa}=\sum_{k=0,\neq 1}^{N_{1}}\alpha _{k}^{\left( 1\right) }\left( 
\begin{array}{c}
\tau _{0k}^{\left( 1\right) } \\ 
\tau _{2k}^{\left( 1\right) }%
\end{array}%
\right) ,\text{ }\vec{\eta}=\sum_{k=0}^{N_{2}}\alpha _{k}^{\left( 2\right)
}\left( 
\begin{array}{c}
\tau _{0k}^{\left( 2\right) } \\ 
\tau _{1k}^{\left( 2\right) }%
\end{array}%
\right) ,  \notag \\
\hat{\tau}_{n}=\sum_{r=0,\neq 1}^{N_{1}}\left( 
\begin{array}{cc}
\tau _{0r}^{\left( 1\right) }\lambda _{r0}^{\left( 1\right) }\left. {}\right.
& \tau _{0r}^{\left( 1\right) }\lambda _{r2}^{\left( 1\right) } \\ 
\tau _{2r}^{\left( 1\right) }\lambda _{r0}^{\left( 1\right) }\left. {}\right.
& \tau _{2r}^{\left( 1\right) }\lambda _{r2}^{\left( 1\right) }%
\end{array}%
\right) ,\text{ }\hat{\tau}_{\pi }=\sum_{r=0}^{N_{2}}\left( 
\begin{array}{cc}
\tau _{0r}^{\left( 2\right) }\lambda _{r0}^{\left( 2\right) }\left. {}\right.
& \tau _{0r}^{\left( 2\right) }\lambda _{r1}^{\left( 2\right) } \\ 
\tau _{1r}^{\left( 2\right) }\lambda _{r0}^{\left( 2\right) }\left. {}\right.
& \tau _{1r}^{\left( 2\right) }\lambda _{r1}^{\left( 2\right) }\text{ }%
\end{array}%
\right) .
\end{gather}%
The transport coefficients of the nonlinear terms in the equation of motion
for $\vec{n}^{\mu }$ are%
\begin{align}
\hat{\delta}_{nn}& =\frac{1}{3}\sum_{r=0,\neq 1}^{N_{1}}\left( 
\begin{array}{cc}
3\tau _{0r}^{\left( 1\right) }\lambda _{r0}^{\left( 1\right) }\left.
{}\right. & 5\tau _{0r}^{\left( 1\right) }\lambda _{r2}^{\left( 1\right) }
\\ 
3\tau _{2r}^{\left( 1\right) }\lambda _{r0}^{\left( 1\right) }\left.
{}\right. & 5\tau _{2r}^{\left( 1\right) }\lambda _{r2}^{\left( 1\right) }%
\end{array}%
\right) \text{ },  \notag \\
\hat{\lambda}_{nn}& =\frac{1}{5}\sum_{r=0,\neq 1}^{N_{1}}\left( 2r+3\right)
\left( 
\begin{array}{cc}
\tau _{0r}^{\left( 1\right) }\lambda _{r0}^{\left( 1\right) }\left. {}\right.
& \tau _{0r}^{\left( 1\right) }\lambda _{r2}^{\left( 1\right) } \\ 
\tau _{2r}^{\left( 1\right) }\lambda _{r0}^{\left( 1\right) }\left. {}\right.
& \tau _{2r}^{\left( 1\right) }\lambda _{r2}^{\left( 1\right) }%
\end{array}%
\right) ,\text{ } \\
\hat{\lambda}_{n\pi }& =\frac{1}{4}\left[ \sum_{r=0}^{N_{2}}\left( 
\begin{array}{cc}
\tau _{00}^{\left( 1\right) }\mathcal{F}_{1r}^{\left( 2\right) }\lambda
_{r0}^{\left( 2\right) }\left. {}\right. & 2\tau _{00}^{\left( 1\right) }%
\mathcal{F}_{1r}^{\left( 2\right) }\lambda _{r1}^{\left( 2\right) } \\ 
\tau _{20}^{\left( 1\right) }\mathcal{F}_{1r}^{\left( 2\right) }\lambda
_{r0}^{\left( 2\right) }\left. {}\right. & 2\tau _{20}^{\left( 1\right) }%
\mathcal{F}_{1r}^{\left( 2\right) }\lambda _{r1}^{\left( 2\right) }%
\end{array}%
\right) +\sum_{r=2}^{N_{1}}\left( 
\begin{array}{cc}
\left( 1-r\right) \tau _{0r}^{\left( 1\right) }\lambda _{r-1,0}^{\left(
2\right) }\left. {}\right. & \left( 2-r\right) \tau _{0r}^{\left( 1\right)
}\lambda _{r-1,1}^{\left( 2\right) } \\ 
\left( 1-r\right) \tau _{2r}^{\left( 1\right) }\lambda _{r-1,0}^{\left(
2\right) }\left. {}\right. & \left( 2-r\right) \tau _{2r}^{\left( 1\right)
}\lambda _{r-1,1}^{\left( 2\right) }%
\end{array}%
\right) \right] , \\
\hat{\tau}_{n\pi }& =-4P_{0}\left[ \sum_{r=2}^{N_{1}}\left( 
\begin{array}{cc}
0\text{ }\left. {}\right. & \tau _{0r}^{\left( 1\right) }\lambda
_{r-1,1}^{\left( 2\right) } \\ 
0\left. {}\right. & \tau _{2r}^{\left( 1\right) }\lambda _{r-1,1}^{\left(
2\right) }%
\end{array}%
\right) +\sum_{r=0}^{N_{2}}\left( 
\begin{array}{cc}
0\text{ }\left. {}\right. & \tau _{00}^{\left( 1\right) }\mathcal{F}%
_{1r}^{\left( 2\right) }\lambda _{r1}^{\left( 2\right) } \\ 
0\left. {}\right. & \tau _{20}^{\left( 1\right) }\mathcal{F}_{1r}^{\left(
2\right) }\lambda _{r1}^{\left( 2\right) }%
\end{array}%
\right) \right] ,\text{ } \\
\hat{\ell}_{n\pi }& =\left[ -\sum_{r=0}^{N_{2}}\left( 
\begin{array}{cc}
\tau _{00}^{\left( 1\right) }\mathcal{F}_{1r}^{\left( 2\right) }\lambda
_{r0}^{\left( 2\right) }\left. {}\right. & \tau _{00}^{\left( 1\right) }%
\mathcal{F}_{1r}^{\left( 2\right) }\lambda _{r1}^{\left( 2\right) } \\ 
\tau _{20}^{\left( 1\right) }\mathcal{F}_{1r}^{\left( 2\right) }\lambda
_{r0}^{\left( 2\right) }\left. {}\right. & \tau _{20}^{\left( 1\right) }%
\mathcal{F}_{1r}^{\left( 2\right) }\lambda _{r1}^{\left( 2\right) }%
\end{array}%
\right) +\frac{\beta _{0}}{4P_{0}}\sum_{r=0,\neq 1}^{N_{1}}\left( 
\begin{array}{cc}
\tau _{0r}^{\left( 1\right) }I_{r+2,1}\left. {}\right. & 0 \\ 
\tau _{2r}^{\left( 1\right) }I_{r+2,1}\left. {}\right. & 0%
\end{array}%
\right) -\sum_{r=2}^{N_{1}}\left( 
\begin{array}{cc}
\tau _{0r}^{\left( 1\right) }\lambda _{r-1,0}^{\left( 2\right) }\left.
{}\right. & \tau _{0r}^{\left( 1\right) }\lambda _{r-1,1}^{\left( 2\right) }
\\ 
\tau _{2r}^{\left( 1\right) }\lambda _{r-1,0}^{\left( 2\right) }\left.
{}\right. & \tau _{2r}^{\left( 1\right) }\lambda _{r-1,1}^{\left( 2\right) }%
\end{array}%
\right) \right] ,
\end{align}%
while those in the equation of motion for $\vec{\pi}^{\mu \nu }$ are 
\begin{align}
\hat{\delta}_{\pi \pi }& =\frac{1}{3}\sum_{r=0}^{N_{2}}\left( 
\begin{array}{cc}
4\tau _{0r}^{\left( 2\right) }\lambda _{r0}^{\left( 2\right) }\left.
{}\right. & 5\tau _{0r}^{\left( 2\right) }\lambda _{r1}^{\left( 2\right) }
\\ 
4\tau _{1r}^{\left( 2\right) }\lambda _{r0}^{\left( 2\right) }\left.
{}\right. & 5\tau _{1r}^{\left( 2\right) }\lambda _{r1}^{\left( 2\right) }%
\end{array}%
\right) ,  \notag \\
\hat{\tau}_{\pi \pi }& =\frac{2}{7}\sum_{r=0}^{N_{2}}\left( 2r+5\right)
\left( 
\begin{array}{cc}
\tau _{0r}^{\left( 2\right) }\lambda _{r0}^{\left( 2\right) }\left. {}\right.
& \tau _{0r}^{\left( 2\right) }\lambda _{r1}^{\left( 2\right) } \\ 
\tau _{1r}^{\left( 2\right) }\lambda _{r0}^{\left( 2\right) }\left. {}\right.
& \tau _{1r}^{\left( 2\right) }\lambda _{r1}^{\left( 2\right) }%
\end{array}%
\right) , \\
\hat{\tau}_{\pi n}& =\frac{1}{5P_{0}}\sum_{r=1}^{N_{2}}\left( 
\begin{array}{cc}
2\tau _{0r}^{\left( 2\right) }\lambda _{r+1,0}^{\left( 1\right) }\left.
{}\right. & 3\tau _{0r}^{\left( 2\right) }\lambda _{r+1,2}^{\left( 1\right) }
\\ 
2\tau _{1r}^{\left( 2\right) }\lambda _{r+1,0}^{\left( 1\right) }\left.
{}\right. & 3\tau _{1r}^{\left( 2\right) }\lambda _{r+1,2}^{\left( 1\right) }%
\end{array}%
\right) , \\
\hat{\ell}_{\pi n}& =\frac{2}{5}\sum_{r=1}^{N_{2}}\left( 
\begin{array}{cc}
\tau _{0r}^{\left( 2\right) }\lambda _{r+1,0}^{\left( 1\right) }\left.
{}\right. & \tau _{0r}^{\left( 2\right) }\lambda _{r+1,2}^{\left( 1\right) }
\\ 
\tau _{1r}^{\left( 2\right) }\lambda _{r+1,0}^{\left( 1\right) }\left.
{}\right. & \tau _{1r}^{\left( 2\right) }\lambda _{r+1,2}^{\left( 1\right) }%
\end{array}%
\right) , \\
\hat{\lambda}_{\pi n}& =-\frac{1}{10}\sum_{r=2}^{N_{2}}\left( 
\begin{array}{cc}
\left( 1+r\right) \tau _{0r}^{\left( 2\right) }\lambda _{r+1,0}^{\left(
1\right) }\left. {}\right. & \tau _{0r}^{\left( 2\right) }\left( r-1\right)
\lambda _{r+1,2}^{\left( 1\right) } \\ 
\left( 1+r\right) \tau _{1r}^{\left( 2\right) }\lambda _{r+1,0}^{\left(
1\right) }\left. {}\right. & \tau _{1r}^{\left( 2\right) }\left( r-1\right)
\lambda _{r+1,2}^{\left( 1\right) }%
\end{array}%
\right) \text{ }.
\end{align}

\end{document}